\documentclass[a4paper]{JHEP3}
\usepackage{amsmath,amsfonts,a4,epsfig,graphicx}
\usepackage{multicol}
\usepackage{float}
\usepackage{slashed} 
\numberwithin{equation}{section}
\def\beq{\begin{equation}}
\def\eeq{\end{equation}}
\def\bea{\begin{eqnarray}}
\def\eea{\end{eqnarray}}

\newcommand{\bit}{\begin{itemize}} 
\newcommand{\eit}{\end{itemize}}

\newcommand{\met}{\ensuremath{\slashed{E}_T}}
\newcommand{\pythia}{{\sc Pythia}}
\newcommand{\delphes}{{\sc Delphes}}

\newcommand{\mgme}{{\sc MadGraph/MadEvent}}

\newcommand{\tauola}{{\sc TAUOLA}}
\newcommand{\madevent}{{\sc MadEvent}}

\newcommand{\kt}{k_T}
\newcommand{\fbinv} {\mbox{\ensuremath{\,\text{fb}^\text{$-$1}}}}
\newcommand{\pbinv} {\mbox{\ensuremath{\,\text{pb}^\text{$-$1}}}}
\catcode`\@=11

\def\@citex[#1]#2{\if@filesw\immediate\write\@auxout{\string\citation{#2}}\fi
  \def\@citea{}\@cite{\@for\@citeb:=#2\do
    {\@citea\def\@citea{,\penalty\@m}\@ifundefined
       {b@\@citeb}{{\bf ?}\@warning
       {Citation `\@citeb' on page \thepage \space undefined}}%
\hbox{\csname b@\@citeb\endcsname}}}{#1}}

\def\citer{\@ifnextchar
[{\@tempswatrue\@citexr}{\@tempswafalse\@citexr[]}}
\def\@citexr[#1]#2{\if@filesw\immediate\write\@auxout{\string\citation{#2}}\fi
  \def\@citea{}\@cite{\@for\@citeb:=#2\do
    {\@citea\def\@citea{--\penalty\@m}\@ifundefined
       {b@\@citeb}{{\bf ?}\@warning
       {Citation `\@citeb' on page \thepage \space undefined}}%
\hbox{\csname b@\@citeb\endcsname}}}{#1}}
\catcode`\@=12

\title{Probing New Physics via $pp \to W^+W^-\to l\nu jj$ at the CERN LHC}

\author{Shuai Liu$^{\, (1)}$, Yajun Mao$^{\, (1)}$, Yong Ban$^{\, (1)}$, Pietro Govoni$^{\, (2),(3),(4)}$, Qiang Li$^{\, (1)}$, Chayanit Asawatangtrakuldee$^{\, (1)}$, Zijun Xu$^{\, (1)}$ \\
$^{\, (1)}$ Department of Physics and State Key
Laboratory of Nuclear Physics and Technology, Peking
University, Beijing, 100871, China \\
$^{\, (2)}$ Milano-Bicocca University, Italy \\
$^{\, (3)}$ INFN Milano-Bicocca, Italy \\
$^{\, (4)}$ European Organization for Nuclear Research 
}

\abstract{TeV scale new Physics, e.g., Large Extra Dimensions or Models with anomalous
triple vector boson couplings, can lead to excesses in various kinematic regions 
on the semi-leptonic productions of $pp \to WW\to l\nu jj$ at the CERN LHC, which, although suffers
from large QCD background compared with the pure leptonic channel,
can benefit from larger production rates and the reconstructable 4-body mass $M_{l\nu jj}$.
We study the search sensitivity through the $l\nu jj$ channel at the 7TeV LHC on relevant new physics, via
probing the hard tails on the reconstructed $M_{l\nu jj}$ and the transverse momentum of $W$-boson ($P_{T\,W}$),
taking into account main backgrounds and including the parton shower and detector simulation effects.
Our results show that with integrated luminosity of $5\fbinv$, the LHC can already discovery or exclude 
a large parameter region of the new physics, e.g., 95\% CL. limit can be set on the Large Extra Dimensions with
a cut-off scale up to 1.5 TeV, and the $WWZ$ anomalous coupling down to, e.g. $|\lambda_Z|\sim 0.1$. Brief results are also given
for the 8TeV LHC.}

\date{\Date}

\keywords{W boson, MC Simulations, Hadron Colliders}

\begin{document}


\section{Introduction}
\label{intr}

The Large Hadron Collider (LHC) had been running successfully during the 2010 and 2011 data taking period
with the c.m.~energy of 7 TeV, accumulating already about $5\fbinv$ data, and will soon be
upgraded to 8 TeV and higher instantaneous luminosity~\cite{lhc}. It has already enriched greatly 
our knowledge on electroweak symmetry breaking (especially the Higgs mechanism), the Standard Model (SM)
precise measurement, SUSY and Extra Dimension physics, etc (see e.g.,  ~\citer{Chatrchyan:2012tx,Vranjes:2012yz}).

Taking the Large Extra Dimensions (ADD)~\cite{ADD} as an example, searches have been performed  
on virtual-graviton channels at HERA~\cite{Adloff:2000dp,Chekanov:2003pw},
LEP~\citer{Acciarri:1999bz,Abbiendi:1999wm}, and the Tevatron~\cite{Abazov:2005tk,Abazov:2008as}.
The most stringent collider limits before the LHC were given by the Tevatron D0 through the dijet~\cite{Abazov:2009mh},
di-photon and di-electron channels~\cite{Abazov:2008as}, which exclude
the ADD cut-off scale $M_s$ up to 1.3-2.1 TeV at 95\% C.L., for 7 to 2 extra dimensions.
Recently, ATLAS and CMS have updated these limits. The 95\% C.L. limits from ATLAS read
as 2.27-3.53 TeV depending on extra dimension number $\delta$ through di-photon search with integrated luminosity
of $2.12\fbinv$~\cite{ATLAS:2011ab}, while CMS gives 2.5 to 3.8 TeV, and 2.3 to 3.8 TeV, through di-photon~\cite{Chatrchyan:2011jx,Chatrchyan:2011fq} and di-lepton searches~\cite{Chatrchyan:2012kc}~\footnote{Searches with a mono-Jet and missing transverse energy via graviton real emission has also been performed at CMS with $36\pbinv$ of data~\cite{Chatrchyan:2011nd}, which however leads to a bit loose limits at 95\% C.L., i.e. 1.68-2.56 TeV for $\delta=6$ to 2.}, respectively, with integrated luminosity of about $2\fbinv$.

On the other hand, searches via gauge boson pair productions have also been proposed in various
references~\citer{Gao:2009eq,YuMing:2011en}. 
The characteristical signal shows as excesses on the transverse momentum ($P_T$) of either vector boson or lepton,
and invariant mass ($M_{VV}$) or transverse mass ($M_T$) of the gauge boson pair. Although the cross sections
in both the SM and ADD are smaller than in the di-photon and di-lepton channels, preliminary analyses~\cite{Gao:2009eq,YuMing:2011en} (without parton shower and detector simulation for $WW$ channel yet) show that 3$\sigma$ sensitivity can be achieved for $M_s$ up to about 2 TeV and 4 TeV, with the integrated luminosity of $1\fbinv$ and $100\fbinv$ at the 7 TeV LHC, respectively. Needless to say that the search via di-boson channel may also be used for combination with other analysis
results to enlarge further the exclusion limits or discovery sensitivity.

In the meantime, di-gauge boson channels are also crucial on probing the Triple Gauge Boson anomalous Coupling (TGC),
which have similar signal type as described above for the ADD and thus can be considered in the same study. 
So far, the tightest cut on the TGC parameter, e.g., $\lambda_Z$ (which is the most interested for us to show as an example
to verify our MC estimation results in this paper, as it is relative independent from other TGC parameters), is from Tevatron D0,
with the resulted 95\% C.L. limits as $-0.075<\lambda_Z<0.093$~\cite{Abazov:2010qn}, 
taking the form factor scale $\Lambda=2$\,TeV, from measurement of $WZ\to l\nu ll$ with $4.1\fbinv$ of data. 
As for $WW$ channel searches which is of our interest, D0 gives 
$-0.10<\lambda_Z<0.11$ through $l\nu jj$ analysis with $1.1\fbinv$ of data~\cite{Abazov:2009tr}, and $-0.14<\lambda_Z<0.18$ through $l\nu l\nu$
analysis with $1\fbinv$ of data~\cite{Abazov:2009hk}. The ATLAS TDR~\cite{Aad:2009wy} also gave an estimate on 
the ability of the 14 TeV LHC, which can reach $-0.028<\lambda_Z<0.024$ with $1\fbinv$ of data by fitting $M_T(WZ)$ in the $WZ$ measurement, and $-0.108<\lambda_Z<0.111$ from $WW\to l\nu l\nu$ channel. Note also recently ATLAS gave constraints as 
$-0.090<\lambda_Z<0.086$ for $\Lambda=3$\,TeV with $1.02\fbinv$ of data by fitting the $P_T$ of the leading lepton in $WW\to l\nu l\nu$ measurement~\cite{atlasww}.

In this paper, we are interested in simulating and exploring the ability of new physics searches via the semi-leptonic
channel of di-$W$ boson productions, $PP \to WW\to l\nu jj$, for both the ADD and TGC searches the LHC.
The semi-leptonic channel, although suffers from larger QCD background compared with the pure leptonic channel,
can benefit from larger production rates and the reconstructable 4-body mass $M_{l\nu jj}$, which can then be used
for shape fitting in data analysis to constrain well the systematics via extrapolation method
from control to signal region. Note the semi-leptonic $l\nu jj$ channel has also been studied extensively before both in MC and experimental analysis, on Higgs search~\citer{Han:1998ma,CMSlvjj} and triple gauge boson anomalous coupling~\cite{Abazov:2009tr}, which shows benefits to increase the search sensitivity either alone or by combining with other channels.

This paper is organized as follows. In
Section~\ref{sab} we list the signal and main backgrounds. In
Section~\ref{simulation} we describe the simulation working line.
In Section~\ref{results} we present numerical results and their
discussion. Finally we conclude in Section~\ref{sec:end}.

\section{Signals and Backgrounds}
\label{sab}

\FIGURE{\label{fey}
\includegraphics[width=14.cm]{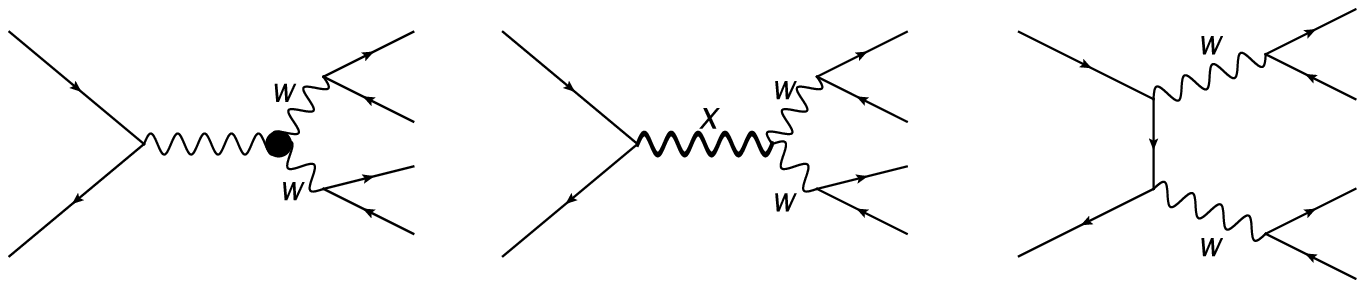}  
\caption{Example Feynman diagrams on $l\nu jj$ productions at the LHC. The bold vertex and line represents the TGC and ADD Kaluza-Klein gravitons.}}

We show in Fig.~\ref{fey} examples of relevant Feynman diagrams for $pp \to W^+W^-\to l\nu jj$ productions at the LHC. Additional contributions from the TGC and Kaluza-Klein (KK) gravitons in the ADD are represented by the bold vertex and line, respectively. Note in the ADD, there also appears $gg$ initial channel, while in the TGC, $pp \to WZ\to l\nu jj$ contributes in addition.

In the ADD model, there are infinite KK modes of gravitons. For virtual graviton channels, the summation over their propagators leads, however, to ultraviolet divergences. This happens because ADD is an effective theory, which is only valid below an effective energy scale. There are several popular ways of parameterizing the LO differential cross sections, including the Han, Lykken, Zhang (HLZ)~\cite{han99:hlz}, the Giudice, Rattazzi, Wells (GRW)~\cite{GRW99:extradim} and the Hewett~\cite{Hewett:1998sn} conventions. In the following we stick to the GRW one which doesn't dependent on $\delta$, in which the summation over the KK graviton propagators can be approximated by
\begin{eqnarray}\label{GRW}
\frac{-1}{\bar{M}^2_{Pl}}\sum_{\vec{n}_\delta}\frac{1}{s-m^2_{\vec{n}_\delta}}=\frac{4\pi}{M^4_s},
\end{eqnarray}
with $\bar{M}_{Pl}$ as the Plank scale and $\vec{n}_\delta$ is a $\delta$-dimension array representing the $n$-th KK mode. In the GRW convention, the above mentioned CMS analyses~\cite{Chatrchyan:2011jx,Chatrchyan:2011fq,Chatrchyan:2012kc} set 95\% C.L. limits on $M_s$ up to about
3 TeV (with some differences depending on the choice of the NLO QCD K factor).
Moreover, we will also present the results with a hard truncation scheme, by setting the cut on the partonic centre-of-mass energy
\begin{eqnarray}\label{TRUC}
\sqrt{\hat{s}}<M_s.
\end{eqnarray}

As for the TGC, we are focusing in this paper as an example on the $\lambda_Z$ term as in the following effective Lagrangian~\cite{anol},
\begin{eqnarray}\label{anola}
{\cal L}^{WWZ}_{\rm eff} \sim -i e \cot\theta_W 
\frac{\lambda_Z}{M_W^2} W^{+\nu}_\mu W^{-\rho}_\nu Z^\mu_\rho,
\end{eqnarray}
where $\theta_W$ is the weak mixing angle, and in the SM one has $\lambda_Z=0$.
Moreover, in order to avoid unitarity violation, we take the following common used dipole form factor
\begin{eqnarray}\label{dipoleanola}
\lambda_Z  \rightarrow  \frac{\lambda_Z}{(1+\hat{s}/\Lambda^2)^2},
\end{eqnarray}
with $\hat{s}$ as the gauge boson pair invariant mass and the parameter $\Lambda$ is fixed to be 2 TeV.

We have used \mgme~\cite{Alwall:2007st,Alwall:2011uj} to deal with the ADD and TGC models. As for the ADD, we have exploited previous implementation of Spin-2 particles~\cite{spin2MG} but now with additional modifications on relevant HELAS subroutines~\cite{helas} to realize the GRW summation convention, i.e. Eq.~(\ref{GRW}). For the TGC, we have used
the FeynRules~\cite{Christensen:2008py}-UFO~\cite{Degrande:2011ua}-ALOHA~\cite{deAquino:2011ub} framework to achieve its implementation within \mgme. The unitary restoration formula Eq.~(\ref{dipoleanola}) is realized by modifying \madevent~\cite{Maltoni:2002qb}.

\FIGURE{\label{partonlevel}
\includegraphics[width=10.cm]{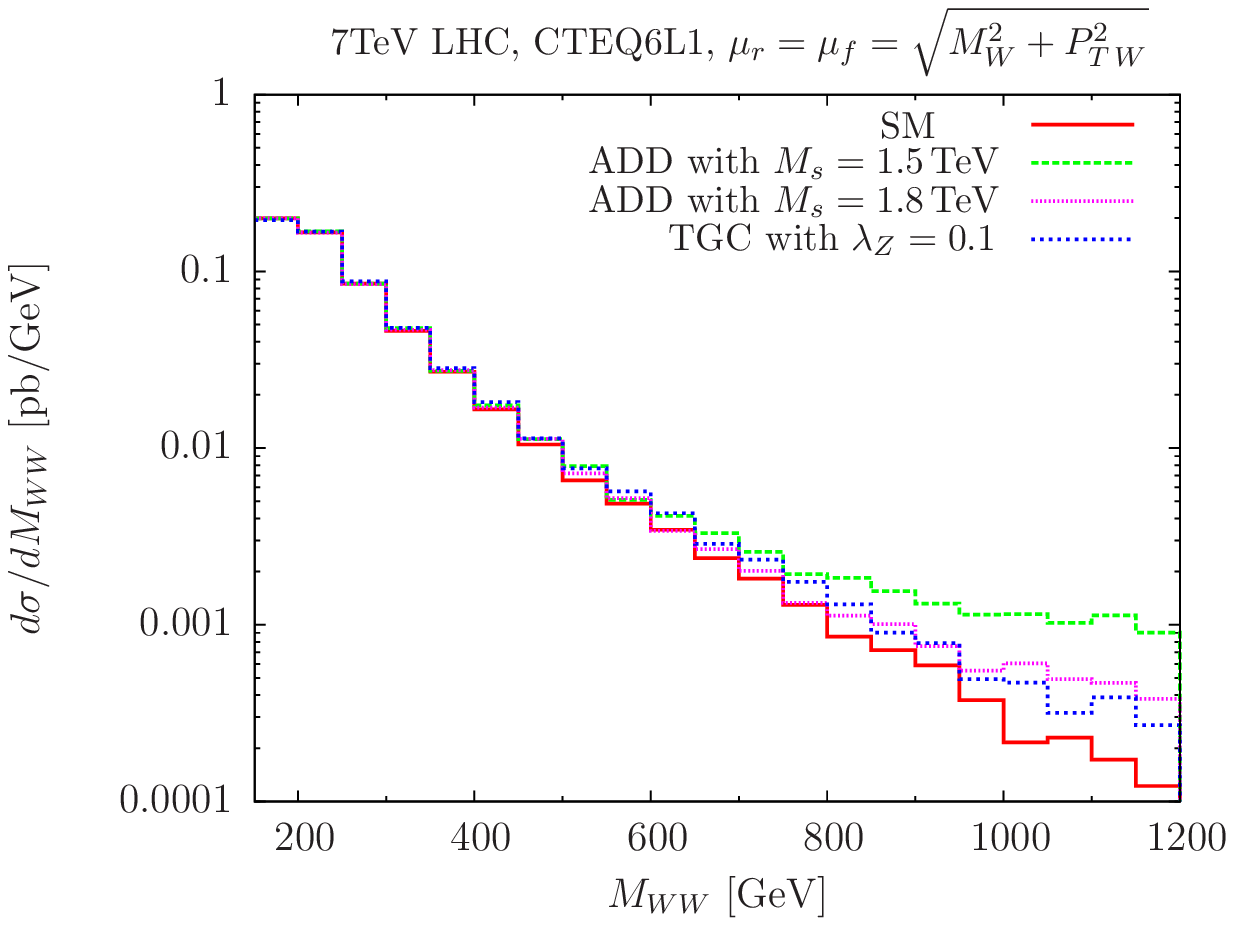}
\includegraphics[width=10.cm]{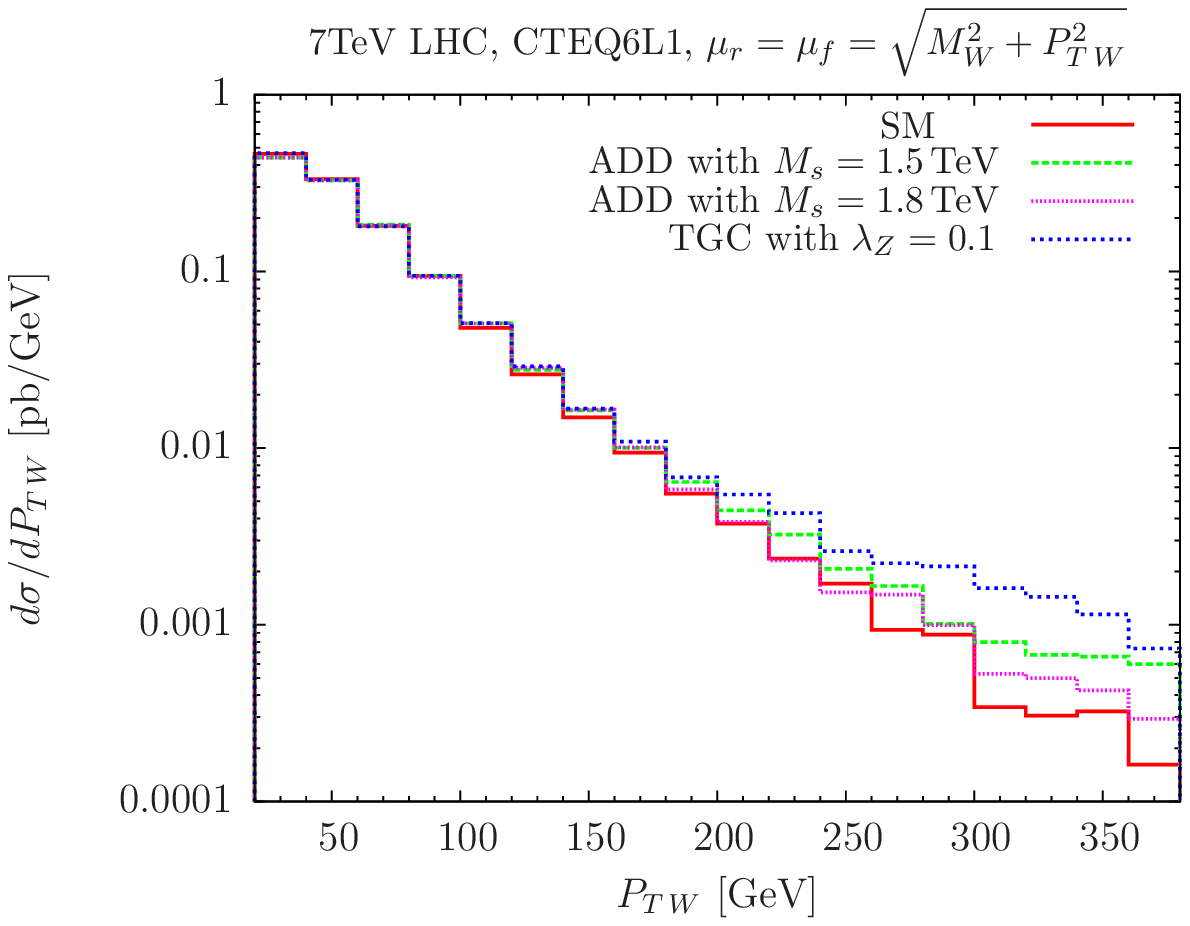}  
\caption{Distributions on $M_{WW}$ and $P_{T\,W}$ for $pp\to W^+W^-$ at the parton level, for the SM, TGC with $\lambda_Z=0.1$ and ADD with $M_s=1.5$ and $1.8$\,TeV.}}

In Fig.~\ref{partonlevel}, we show the $M_{WW}$ and $P_{T\,W}$ differential distributions for $pp\to W^+W^-$ (without $W$ decay) at the parton level, for the SM, TGC with $\lambda_Z=0.1$ and ADD with $M_s=1.5$ and $1.8$\,TeV, with the total cross sections as 28.75, 29.34, 30.63 and 29.3pb, respectively. Those results are for the 7 TeV LHC and the renormalization/factorization scales are set to $\mu_r=\mu_f=\sqrt{P^2_{T\,W}+M^2_W}$. One can see that both ADD and TGC lead to excesses on the hard tails of the $M_{WW}$ and $P_{T\,W}$ distributions: the ADD excesses appear from $M_{WW}\gtrsim 500-700$\,GeV and $P^T_{W}\gtrsim 200-250$\,GeV for $M_s=1.5$ and $1.8$\,TeV, yet the TGC excesses appear a bit more early from $M_{WW}\gtrsim 400$\,GeV and $P^T_{W}\gtrsim 100$\,GeV and seem more globally. One in principle can use the excesses to extract signal out from the backgrounds to a large extent as one can, but to do that it needs a more realistic and complete simulation, taking into account acceptance efficiency, parton shower and detector simulation effects, which will be discussed and shown in detail in the following.

The characterical signal we are interested in contains two well identified leptons (electrons and muons) in association with large \met\,. The main backgrounds are listed as following
\begin{itemize}{
\item (1) $WW\to l\nu + jj$, 
\item (2) $WZ \to l\nu + jj$, 
\item (3) $WZ, ZZ \to jj + ll$ with one lepton misidentified,
\item (4) $W(\to l \nu)+2$-jets, which is the dominant background in our case,
\item (5) $Z(\to ll)+2$-jets with one lepton misidentified,
\item (6) $t\bar{t} \to l\nu jj +b\bar{b}$, 
\item (7) $tW \to l\nu jj +b$,
\item (8) $tj \to l^+\nu j+b$.
}\end{itemize}
Note we have included $\tau$ lepton in the $W$ and $Z$ decay products, which decays into $e, \mu$ at the ratio of about 35\% and is handled with \tauola~\cite{tauola}.
For (4), we have also compared the results with the one of matrix elements for $W+1,\,2,\,3$ partons matched 
via \pythia\,6~\cite{Sjostrand:2003wg} in the $\kt$-jet MLM scheme~\cite{Alwall:2007fs} implemented in \mgme. The matched results agree well with the $W+2$-jets one in general on shapes for e.g. reconstructed 
$M_{l\nu j j}$, yet with an enhancement of a factor $k\sim 1.5$ at the hard tail. For simplicity, we still use the  
$W+2$-jets sample in our study as it takes much less computing time and size to reach higher statistics. Moreover, we have omitted the QCD multi-jet backgrounds with fake lepton, which is important mostly for electron channel~\cite{ATLAS:2011ae,CMSlvjj} and hard to simulate due to instrumental effects, but fortunately is much smaller than the dominant $W+2$-jets backgrounds~\cite{ATLAS:2011ae,CMSlvjj}. 

\section{Simulation Framework}
\label{simulation}

As mentioned above, we use \mgme\, for hard process generation, with the default renomalization and factorization scales chosen as 
the transverse mass of the core process. The NLO or NNLO QCD K factors are included later for normalization, taken from MCFM~\cite{mcfm} or relevant literature. In more detail, from MCFM, we assign a K factor of 1.52 for $WW$ productions~\cite{Campbell:2011bn}, 1.67 for $WZ$~\cite{Campbell:2011bn}, 1.0 for $W/Z+2$-jets~\cite{Campbell:2002tg,Campbell:2003hd}, $1.02$ for $tW$~\cite{Campbell:2005bb}, $0.8$ for $tj$~\cite{Campbell:2009ss}. According to Ref.~\cite{Cacciari:2011hy}, we assign a K factor of $1.52$ for $t\bar{t}$ production to normalize the \mgme\, LO result to the up-to-date NNLO one. As for the $WW$ productions in the ADD model, we set the K factor the same as in the $WW$ channel (note however it may be larger than the SM K factor according to Ref.~\cite{Agarwal:2010sp,Agarwal:2010sn}).

The generated unweighted events at parton level are then interfaced with \pythia\,6 for parton showering and hadronization. The multiple interaction option is also switched on. The detector simulation is realized with the help of
\delphes V2.0~\cite{Ovyn:2009tx}, where we focus on the CMS detector at the LHC. Finally, the sample analysis are performed with the program package ExRootAnalysis~\cite{ExRootAnalysis} and ROOT~\cite{root}.

\section{Numerical Results}
\label{results}
 
We chose the following preselection cuts to generate unweighted events at parton level
with \mgme\, to interface later with \pythia\, and \delphes\,, 
\begin{itemize}{
\item $P_{T\,j} \geq 20\,$GeV, $|\eta_j|<5$ and $R_{jj}>0.3$.  
\item $P_{T\,l} \geq 10\,$GeV, and $|\eta_l|<3$.  
\item $R_{jl}>0.3$.  
}\end{itemize}
for the signals and backgrounds listed in Sec.~\ref{sab}.
Note, however, for the backgrounds (3) $WZ, ZZ \to jj + ll$  and (5) $Z(\to ll)+2$-jets, we don't require any of the above cuts in order not to make bias on the misidentified leptons.

Tighter cuts are then imposed first on the reconstruction objects in the \delphes\, settings cards,
\begin{itemize}{
\item $P_{T\,e,\mu} \geq 30\,$GeV, and $|\eta_{e,\mu}|<2.4$.  
\item Jets are clustered according to the anti$-k_t$ algorithm with a cone radius $\Delta R = 0.6$. 
Moreover, $P_{T,j}>30$\,GeV and $|\eta_{j}|<5$ are required.
}\end{itemize}

Other high level cuts are set in the analysis steps as following
\begin{itemize}{
\item (A) 1 and only 1 lepton passing the above requirements.  
\item (B) 2 or 3 jets.  
\item (C) Choose as $W$-products the pair of jets with invariant mass closer to the $W$ mass, and ask their $|M_{jj}-M_W|<40$\,GeV. 
\item (D) $R_{lj}>0.4$ 
\item (E) \met\,$>30$\,GeV. 
\item (F) $M^W_T>30$\,GeV. 
\item (G) B-jet veto with the efficiency set as 40\% in the \delphes\, detector card.  

}\end{itemize}

In Table.~\ref{cutf}, we list the event numbers after each step of the analysis cuts, for an 
integrated luminosity of $5\fbinv$ at the 7 TeV LHC, for both the signal and background processes,
where we take the ADD model with $M_s=1.5$\,TeV as an example for the signal. The dominant backgrounds
so far are the $W+2$-jets and $t\bar{t}$. The signal excess $S$ reads as $(\rm{ADD}-\rm{WW})$ and the resulting
$S/\sqrt{B}$ is only about 0.2. Further cuts must be exploited to enlarge the sensitivity, where hints 
lie in Fig.~\ref{partonlevel}.
 
\begin{table*}[h!] 
\begin{center} 
\begin{tabular}{c||c|c|c|c|c|c|c|c|c} 
cut & $WW$ &  $WZ(l\nu jj)$ & $jjll$ & $Wjj$  & $Zjj$ &  $t\bar t$ & tw & tj & ADD(1.5TeV) $\:$ \\  
\hline 
(A)  & 22265  & 3866 & 1451 & 2824837  & 257506  & 443210  &  6612  & 15836  &  23977 \\\hline 
(B)  & 13900  & 2637 & 879  & 1673215  & 149344  & 147350  &  3793  & 11774  &  14989  \\\hline 
(C)  & 10666  & 1995 & 663  & 912832   & 82501   & 104754  &  2805  &  5987  &  10911   \\\hline 
(D)  & 4117   &  894 & 283  & 298093   & 28110   & 70994   &  1672  &  1666  &  4270    \\\hline 
(E)  & 2654   &  596 & 123  & 183820   & 12129   & 53903   &  1191  &  1222  &  2789   \\\hline 
(F)  & 2292   &  512 & 96   & 158679   & 9868    & 42494   &  947   &  1047  &  2374   \\\hline 
(G)  & 2168   &  489 & 91   & 153022   & 9558    & 23834   &  623   &  614   &  2255   \\\hline 
\end{tabular} 
\caption{Cut flow at the LHC with $\sqrt{s}=7$ TeV and an 
  integrated luminosity of $5\fbinv$. 
\label{cutf}} 
\end{center} 
\end{table*} 

In the following, we further vary additional cuts on following variables to optimize the signal background significance:
\begin{itemize}{
\item $\phi_{l,j}$,  
\item $\Delta\eta_{jj}$, where the two jets correspond to the ones got from above (C).
\item $P_{T\,W}$ for the leptonic decayed $W$,
\item $M_{l\nu jj}$.~\footnote{The neutrino momentum is extracted by imposing the invariant mass of the lepton and neutrino to be $m_W$. In case there are two solutions for neutrino momentum, we choose
the one with smaller $p_z$.}
}\end{itemize}
We note the cuts on $P_{T\,W}$ and $M_{l\nu jj}$ are inspired from the parton level results as shown in Fig.~\ref{partonlevel}, as ADD and TGC lead to excesses at their hard tails. The cuts on $\phi_{l,j}$ and $\Delta\eta_{jj}$ 
are considered, as the signals tend to have two close jets back to the lepton, especially in the highly boosted case, while for the QCD background it is more flat.

In Figs.~\ref{m4} and \ref{ptw}, we show the distributions on $M_{l\nu j j}$ and $P_{T\,W}$ at the 7 TeV LHC,
for various backgrounds and the signals in the ADD model, i.e. the ADD Model with $M_s$ as 1.5 TeV, 1.8 TeV, and the Truncated ADD with 1.5TeV. Similar as what we have already seen from the parton level results in Fig.~\ref{partonlevel}, the ADD signals tend to have hard tail. With further cuts as
$M_{l\nu jj}>0.9$\,TeV, $P_{T\,W}>200$\,GeV, $\phi_{l,j}>0.8$, $\Delta\eta_{jj}<2.5$ and the minimum of $|M_{jj}-M_W|<20$\,GeV,
we get $B=19.1$, while for 1.5 TeV ADD, we have $S=9.4$ and $S/\sqrt{B}=2.15$, thus with about $4.15\fbinv$ it is
already enough to reach 95\% C.L. exclusion limit. 
For 1.5 TeV truncated ADD, we have $S=6.0$ and $S/\sqrt{B}=1.37$. For $M_s$ at 1.8TeV and above, the significance can only reach about $0.4$. To get close to 95\% limit, one needs over $100\fbinv$ of data.
Note although the sensitivities via $l\nu jj$ channel appear to be lower than the above mentioned CMS ones~\cite{Chatrchyan:2011jx,Chatrchyan:2011fq,Chatrchyan:2012kc} which set 95\% C.L. limits on $M_s$ up to about 3 TeV in the GRW convention, the $l\nu jj$ channel can still play a complementary role for further confirmation or improvement through combination.

\FIGURE{\label{m4}
\includegraphics[width=11.cm]{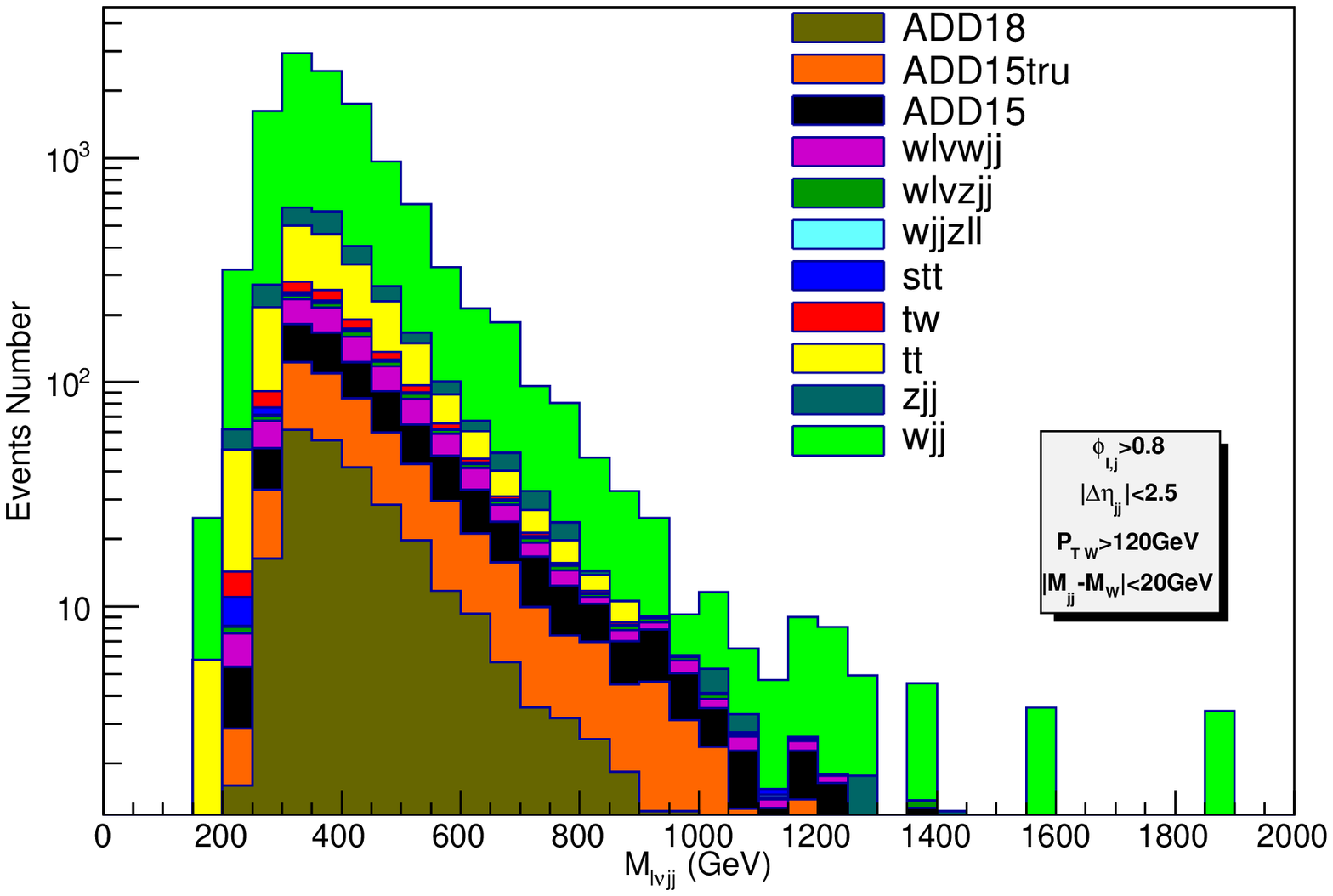}
\caption{$M_{l\nu jj}$ distributions for various backgrounds and the signals in the ADD model, i.e. the ADD Model with $M_s$ as 1.5 TeV, 1.8 TeV, and the Truncated ADD with 1.5TeV, at the LHC with $\sqrt{s}=7$ TeV and an 
integrated luminosity of $5\,\rm{fb}^{-1}$.}}

\FIGURE{\label{ptw}
\includegraphics[width=11.cm]{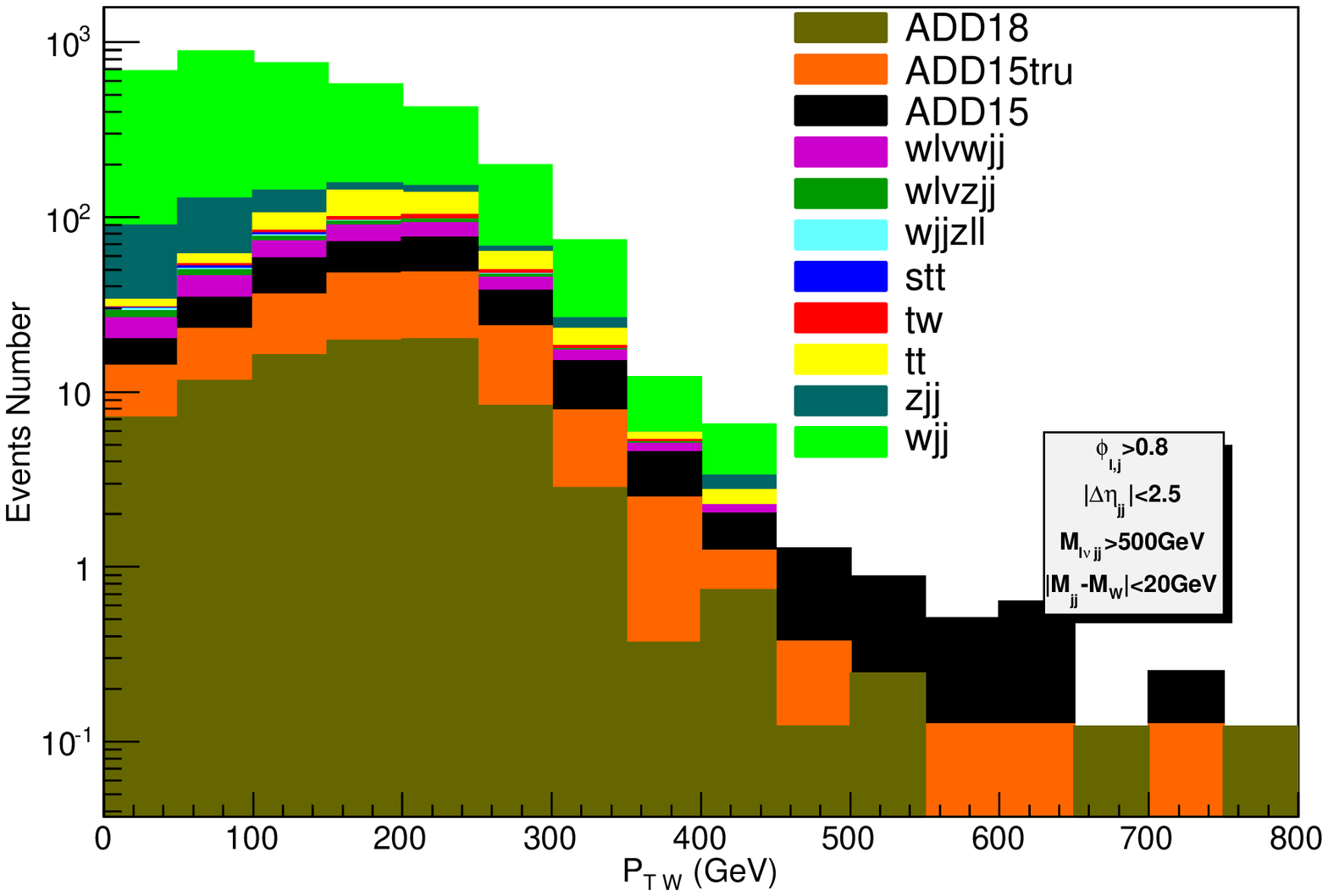}
\caption{$P_{T\,W}$ distributions for various backgrounds and the signals in the ADD model, at the LHC with $\sqrt{s}=7$ TeV and an integrated luminosity of $5\fbinv$. }}

In Figs.~\ref{m4b} and ~\ref{ptwb}, we show similar results as in Figs.~\ref{m4} and ~\ref{ptw}, but now for the TGC signals. With cuts as $M_{l\nu jj}>500$\,GeV, $P_{T\,W}>220$\,GeV, $\phi_{l,j}>0.8$, $\Delta\eta_{jj}<1.5$ and $0<M_{jj}-M_W<25$\,GeV (note this is chosen to optimise the sensitivity through changing the relative weight between $WW$ and $WZ$ channel),
we have $B=589.3$, while for $\lambda_Z=0.1$ TGC, we have $S=47$ and thus $S/\sqrt{B}=1.94$. 
For $\lambda_Z=0.06$ TGC, we have $S=15$ and $S/\sqrt{B}=0.62$. We note these results are more or less near
the ones in the ATLAS TDR~\cite{Aad:2009wy}, where 95\% C.L. limit of $-0.108<\lambda_Z<0.111$ can be got
from $WW$ channel with $1\fbinv$ of data at the 14 TeV LHC, where the cross section
for e.g. $WW$ processes gets increased by a factor of 3-4 compared with the 7 TeV LHC.

\FIGURE{\label{m4b}
\includegraphics[width=11.cm]{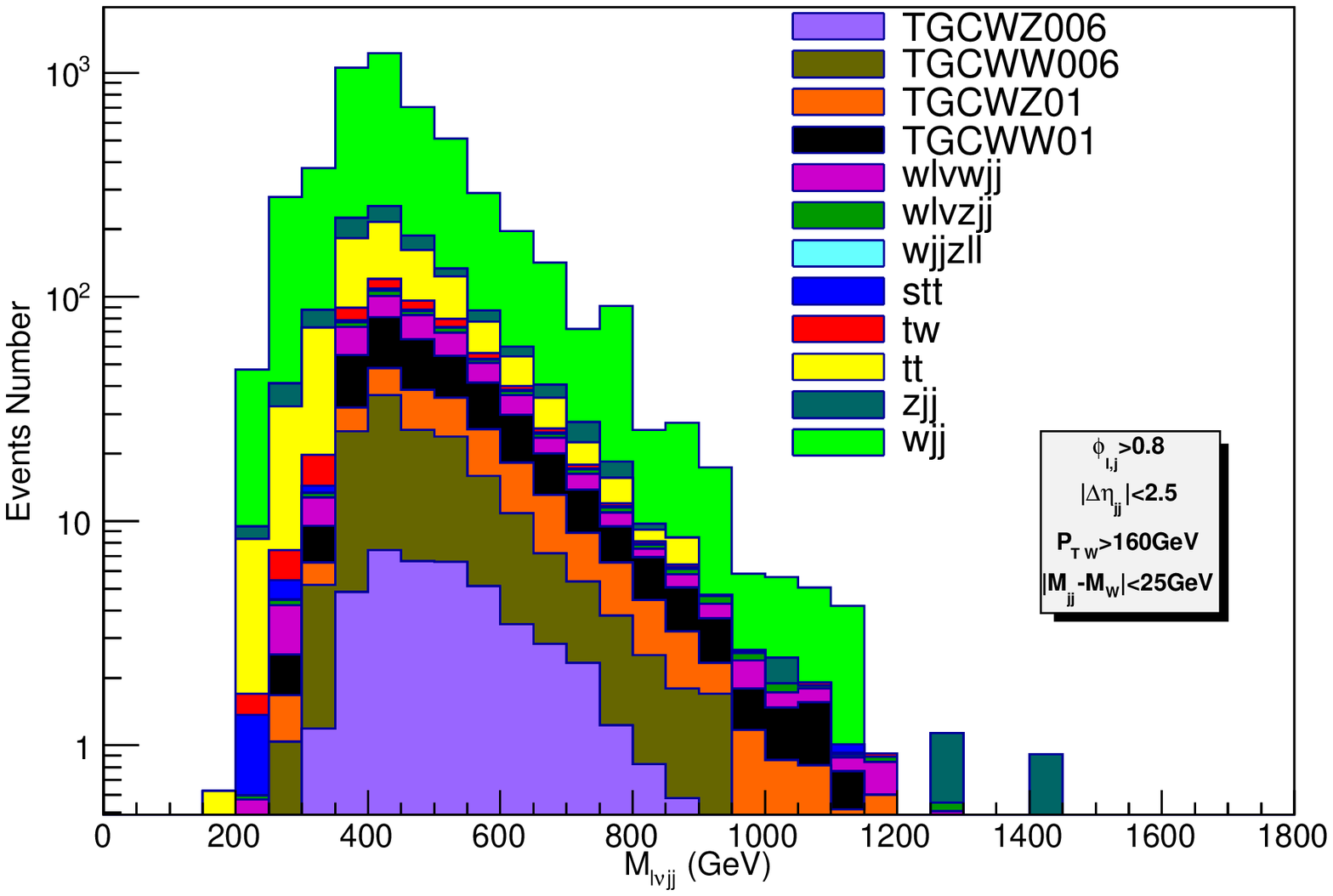}
\caption{$M_{l\nu jj}$ distributions for various backgrounds and the signals in the TGC Model with $\lambda_Z=0.06$ and $0.1$, at the LHC with $\sqrt{s}=7$ TeV and an integrated luminosity of $5\fbinv$.}}

\FIGURE{\label{ptwb}
\includegraphics[width=11.cm]{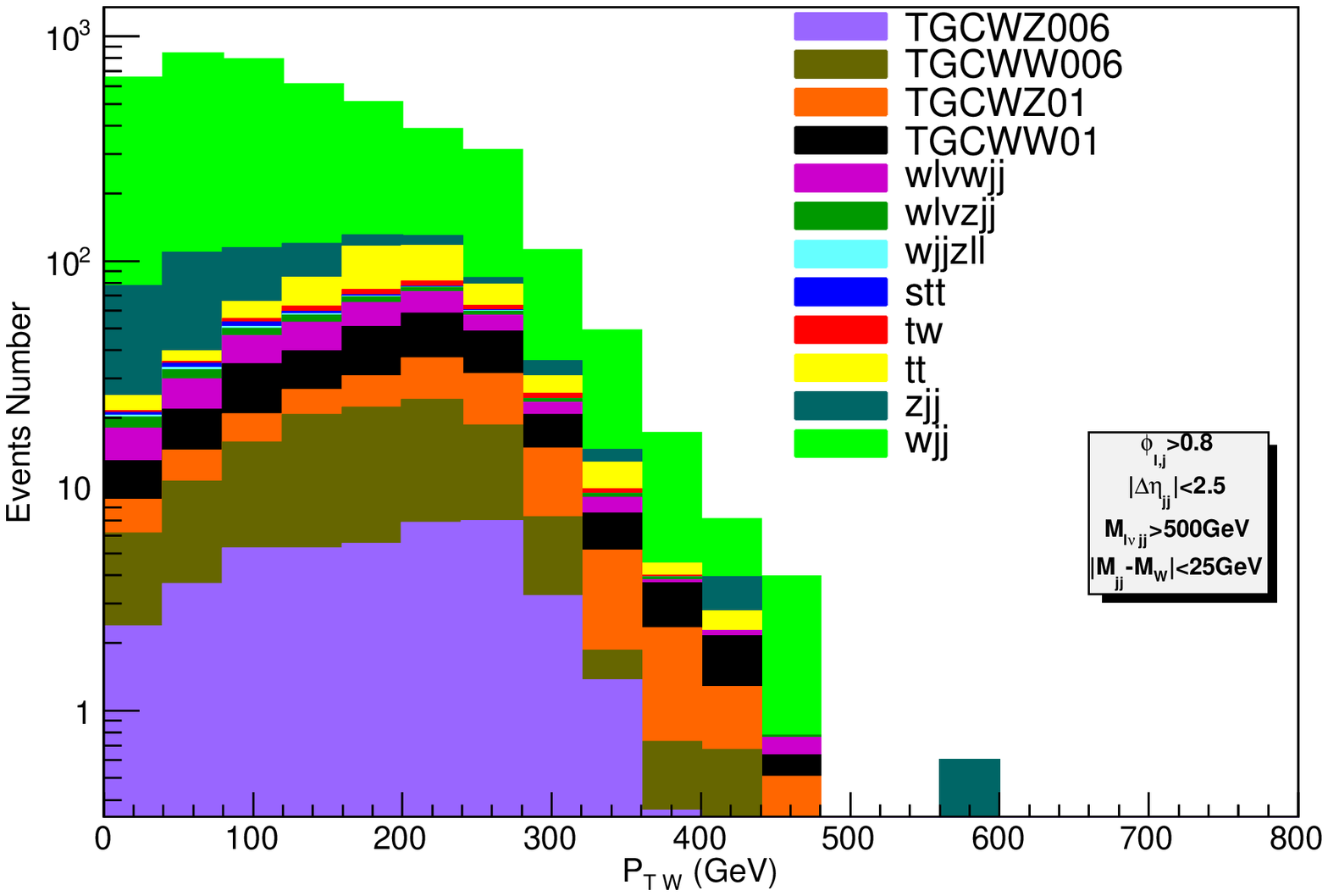}
\caption{$P_{T\,W}$ distributions for various backgrounds and the signals in the TGC Model with $\lambda_Z=0.06$ and $0.1$, at the LHC with $\sqrt{s}=7$ TeV and an integrated luminosity of $5\fbinv$}}

In Table.~\ref{s78}, we show the results of $S/\sqrt{B}$ for searching ADD with $M_s=1.5$\,TeV and $1.8$\,TeV, with or without truncation, at the LHC with $\sqrt{s}=7$\,TeV and $8$\,TeV, respectively. We also show the uncertainty of the significance, for simplicity only by enlarging or decreasing the background by a factor of 2. However, we note here that the systematics can get controlled much better in relevant experimental analysis via data driven method. While evaluating the numbers in the table, we set $\phi_{l,j}>0.8$, $\Delta\eta_{jj}<2.5$ and the minimum of $|M_{jj}-M_W|<20$\,GeV. For ADD with $M_s=1.5$TeV, with or without truncation (Eq.~\ref{TRUC}), we take further 
$M_{l\nu jj}>0.9$\,TeV and $P_{T\,W}>200$\,GeV. 
At the 8 TeV LHC, the sensitivities gets larger compared with the 7 TeV LHC, 
especially for larger $M_s$ case, which can reach $0.61$ and $0.42$ for e.g. $M_s=1.8$\,TeV, with or without truncation. 

\begin{table*}[h!] 
\begin{center} 
\begin{tabular}{c||c|c|c|c|c|c|c|c|c} 
ADD & 1.5TeV &  1.5TeV Truncated & 1.8TeV &  1.8TeV Truncated
\\  
\hline 
7TeV LHC  & $2.15^{+0.89}_{-0.63}$  & $1.37^{+0.57}_{-0.40}$ & $0.37^{+0.12}_{-0.11}$  & $0.34^{+0.14}_{-0.10}$
\\\hline 
8TeV LHC  & $2.22^{+0.92}_{-0.65}$  & $1.78^{+0.74}_{-0.52}$ & $0.61^{+0.25}_{-0.18}$  & $0.42^{+0.17}_{-0.11}$  \\\hline  
\end{tabular} 
\caption{Sensitivity on ADD searches via $WW\to l\nu jj$ channel at the LHC with $\sqrt{s}=7$ TeV and $8$ TeV, respectively, and an 
integrated luminosity of $5\fbinv$. The corresponding uncertainty are also shown, for simplicity just by enlarging or decreasing the background by a factor of 2.
\label{s78}} 
\end{center} 
\end{table*} 

\section{Summary}
\label{sec:end}
In summary, we have presented here the first MC simulation study on searching the Large Extra Dimensions via the $l\nu jj$ channel, taking into account the parton shower and detector simulation effects. We have also updated the results for probing triple gauge boson anomalous coupling. Our results show that with only $5\fbinv$ of data, 
the 7TeV LHC is sensitive to the Large Extra Dimensions with energy scale at $M_s \sim 1.5$\,TeV and the triple gauge boson anomalous coupling, e.g. $|\lambda_Z|\sim 0.1$. At the 8 TeV LHC, the sensitivities get further enhanced. 
For larger $M_s (>1.8)$\,TeV  case, the LHC can only achieve the sensitivity at about $0.4 \sim 0.5$ level, which may be used for combining with other channels, e.g. the leptonic decay one as $l\nu l\nu$, and will be shown in our further works. 

\acknowledgments
This work is supported in part by the National Natural Science Foundation of China, under
Grants No. 10721063, No. 10975004 and No. 10635030.


\appendix


\end{document}